\documentclass[runningheads]{llncs}
\usepackage{graphicx}
\usepackage{amsmath,amssymb,amsfonts}
\usepackage{subfigure}
\usepackage{multirow}
\begin{document}

\title{Profile Matching Across Online Social Networks}

\author{Anisa Halimi \inst{1} \and Erman Ayday \inst{1,2}}
 \institute{Case Western Reserve University, Cleveland, OH, USA
 \and Bilkent University, Turkey \\
 \email{\{anisa.halimi,erman.ayday\}@case.edu}}
\maketitle              

\begin{abstract}
In this work, we study the privacy risk due to profile matching across online social networks (OSNs), in which anonymous profiles of OSN users are matched to their real identities using auxiliary information about them. We consider different attributes that are publicly shared by users. Such attributes include both strong identifiers such as user name and weak identifiers such as interest or sentiment variation between different posts of a user in different platforms. We study the effect of using different combinations of these attributes to profile matching in order to show the privacy threat in an extensive way. The proposed framework mainly relies on machine learning techniques and optimization algorithms. We evaluate the proposed framework on three datasets (Twitter - Foursquare, Google+ - Twitter, and Flickr) and show how profiles of the users in different OSNs can be matched with high probability by using the publicly shared attributes and/or the underlying graphical structure of the OSNs. We also show that the proposed framework notably provides higher precision values compared to state-of-the-art that relies on machine learning techniques. We believe that this work will be a valuable step to build a tool for the OSN users to understand their privacy risks due to their public sharings.

\keywords{social networks \and profile matching \and deanonymization.}
\end{abstract}

\section{Introduction}
An online social network (OSN) is a platform, in which, individuals share vast amount of information about themselves such as their social and professional life, hobbies, diseases, friends, and opinions. Via OSNs, people also get in touch with other people that share similar interests or that they already know in real-life~\cite{ellison2007social}. With the widespread availability of the Internet, especially via mobile devices, OSNs have been a part of our lives more than ever. Most individuals have multiple OSN profiles for different purposes. Furthermore, each OSN offers different services via different frameworks, leading individuals share different types of information~\cite{debnath2008feature}. Also, in some OSNs, users reveal their real identities (e.g., to find old friends), while in some OSNs, users prefer to remain anonymous (especially in OSNs in which users share sensitive information about themselves). 

It is trivial to link profiles of individuals across different OSNs in which they share their real identities. However, such profile matching is both nontrivial and sometimes undesired if individuals do not reveal their real identities in some OSNs. While profile matching is useful for online service providers to build complete profiles of individuals (e.g., to provide better personalized advertisement), it also has serious privacy concerns. If an attacker can link anonymous profiles of individuals to their real identities (via their other OSN accounts in which they share their real identity), they can obtain privacy-sensitive information about individuals that is not intended to be linked to their real identities. Such sensitive information can then be used against the individuals for discrimination or blackmailing. Thus, it is very important to quantify and show to the OSN users the extent of this privacy risk.

Some OSNs can be characterized by their graphical structures (i.e., connections between their users). The graphical structures of some popular OSNs show strong resemblance to social connections of individuals in real-life (e.g., Facebook). Therefore, it is natural to expect that the graphical structures of such OSNs will be similar to each other as well. Existing work shows that this similarity in graphical structure (along with some background information) can be utilized to link accounts of individuals from different OSNs~\cite{narayan2}. However, without sufficient background information, just using graphical structure for profile matching becomes computationally infeasible. On the other hand, some OSNs or online platforms either do not have a graphical structure at all (e.g., forums) or their graphical structure does not resemble the real-life connections of the individuals. However, this does not mean that users of such OSNs are protected against profile matching (or deanonymization). In these types of OSNs, an attacker can utilize the attributes of the users (i.e., types of information that are shared by the users) across different OSNs for deanonymization.

In this work, we quantify and show the risk of profile matching in OSNs by considering both the graphical structure and other attributes of the users. We show the threat between an auxiliary OSN (in which users share their real identities) and a target OSN (in which users prefer to make anonymous sharings). The proposed framework matches user profiles across multiple OSNs by using machine learning and optimization techniques. We mainly focus on two types of attacks (i) targeted attack, in which the attacker selects a set of victims from the target OSN and wants to determine the profiles of the victims in the auxiliary OSN, and (ii) global attack, in which the attacker wants to deanonymize the profiles of all the users that are in the anonymous OSN (assuming they have accounts in the auxiliary OSN). Our results show that by using different machine learning (logistic regression and support vector machine) and optimization techniques, individuals' profiles can be matched with more than $70\%$ accuracy (depending on the set of attributes used for profile matching). We also study the effect of different types of attributes (i.e., strong identifiers and weak identifiers) to the profile matching risk. The main contributions of this work can be summarized as follows:
\vspace{-2pt}
\begin{itemize}
	\item We develop a profile matching framework across OSNs by using various publicly shared attributes of the users and the graphical structure on the OSNs. Using this framework, we show how the privacy risk can be quantified accurately.
	\item We study the effect of different sets of publicly shared attributes to profile matching. In particular, we show how strong identifiers (such as user name and location) and weak identifiers (such as activity patterns across OSNs, interests, or sentiment) of the users help the attacker.
	\item We evaluate the proposed attack on four different social networks.
	\item We show that our profile matching algorithm provides significantly higher precision and a comparable recall to the state-of-the-art.
\end{itemize}

The rest of the paper is organized as follows. In the next section, we summarize the related work and the main differences of this work from the existing works in the area. In Section~\ref{sec:threat}, we discuss the threat model. In Section~\ref{sec:model}, we provide the details of the proposed framework. In Section~\ref{sec:evaluation} we show the results of the proposed framework by using real data.  Finally, in Section~\ref{sec:conclusion}, we discuss the future work and conclude the paper. 

\section{Related Work}\label{sec:related_work}

We review two primary lines of related research: (i) deanonymization based on network structure and (ii) profile matching using public data.

\noindent\textbf{Graph Deanonymization:} In the literature, most works focus on profile matching (or deanonymization) by using structural information that mainly relies on the network structure of OSNs. Narayanan and Shmatikov propose a framework for analyzing privacy and anonymity in social networks and a deanonymization (DA) algorithm that is purely based on network topology~\cite{narayan2}. Another approach by Wondracek et al. uses group membership found on social networks to identify users~\cite{wondracek}. Nilizadeh et al. propose a community-level DA attack~\cite{nilizadeh} by extending the work in~\cite{narayan2}. Unlike previous attacks, Pedarsani et al. propose a seed-free DA attack~\cite{pedarsani}. It is a Bayesian-based model for graph DA which uses degrees and distances to other nodes as each node's fingerprint. Sharad and Danezis propose an automated approach to re-identify users in anonymized social networks~\cite{sharad}. Ji et al. propose a secure graph data sharing/publishing system~\cite{secgraph} in which they implement and evaluate graph data anonymization algorithms, data utility metrics, and modern structure-based deanonymization attacks. 

\noindent\textbf{Profile Matching Using Public Attributes:} It has been shown that by leveraging public information in users' profiles (such as user name, profile photo, description, location, and number of friends) users in different OSNs can be linked to each other. Most works apply different classifiers to the feature vectors to distinguish between matching and non-matching profiles. In Section~\ref{sec:results}, we simulated some of these approaches and we show that our proposed framework provides higher precision compared to them. The attributes used for profile matching vary from one work to another. Shu et al. provide a comprehensive review of state-of-the-art profile matching algorithms~\cite{shu}. Iofciu et al. use only user names and their tags (separately or together) to link different users~\cite{iofciu}. Nunes et al. apply different classifiers to the feature vectors consisting of user name, posts, and sets of friends similarities~\cite{nunes}. Vosecky et al. only use nick name, email, and date of birth to link different  users~\cite{vosecky}. Malhotra et al. use user name, name, description, location, profile photo, and number of connections~\cite{malhotra}. On the other hand, Liu et al. propose a method to match user profiles across multiple communities by using the rareness and commonness of user names~\cite{liu2013}. Zafarani et al. analyze the behaviour patterns of the users, the language used, and the writing style to link users across social media sites~\cite{zafarani}. To evaluate the quality of different user attributes in profile matching, Goga et al. identify four properties: availability, consistency, non-impersonability, and discriminability~\cite{goga2015reliability}. Liu et al. propose a framework called HYDRA that uses both structural and unstructural information to match profiles~\cite{liu2014hydra}. 
Wang et al.~\cite{wang2019user} propose a method that leverages both structural and content information (extracted topics) in a unified way. Zhou et al.~\cite{zhou2019translink} analyze the connections of the users and their behaviours. 

\noindent\textbf{Contribution of this work:} Previous works show that there exists a non-negligible risk of matching user profiles. As the amount of information provided on social networks increases, this risk also increases. However, existing methods mostly focus on accuracy, and hence they provide high false positive rates. They do not use precision and recall (which are shown to be more reliable evaluation metrics~\cite{goga2015reliability}) for evaluation. In this work, we propose a framework that achieves significantly higher precision and a comparable recall to previous works for both structured and unstructured OSNs. Moreover, we consider a wider spectrum of attributes and extensively analyze the effect of weak identifiers to the profile matching scheme. 

\section{Threat Model}\label{sec:threat}

For simplicity, we consider two OSNs to describe the threat: (i) $A$, the auxiliary OSN that includes the profiles of individuals with their identifiers and (ii) $T$, the target OSN that includes anonymous profiles of individuals. In general, the attacker knows the identity of the individuals from OSN $A$ and depending on the type of the attack, they want to determine the real identities of the user(s) in OSN $T$ by only using the public attributes of the users (i.e., information that is publicly shared by the users). The attacker can be a part (user) of both OSNs and they can collect publicly available data from both OSNs (e.g., via crawling). We assume that the attacker is not an insider in $T$. That is, the attacker cannot use the IP address, access patterns, or sign up information of the victim for profile matching (or deanonymization).

We consider two different attacks (i) targeted attack, and (ii) global attack. In the targeted attack, the attacker wants to deanonymize the anonymous profile of a victim (or a set of victims) in OSN $T$, using the unanonymized profile of the same victim in OSN $A$. In the global attack, the attacker's goal is to deanonymize the anonymous profiles of all individuals in $T$ by using the information in $A$. An attacker can select either attack model based on their goals and resources.

\section{Proposed Model}\label{sec:model}

\begin{figure}[ht]
	\centering
	\includegraphics[scale=0.46]{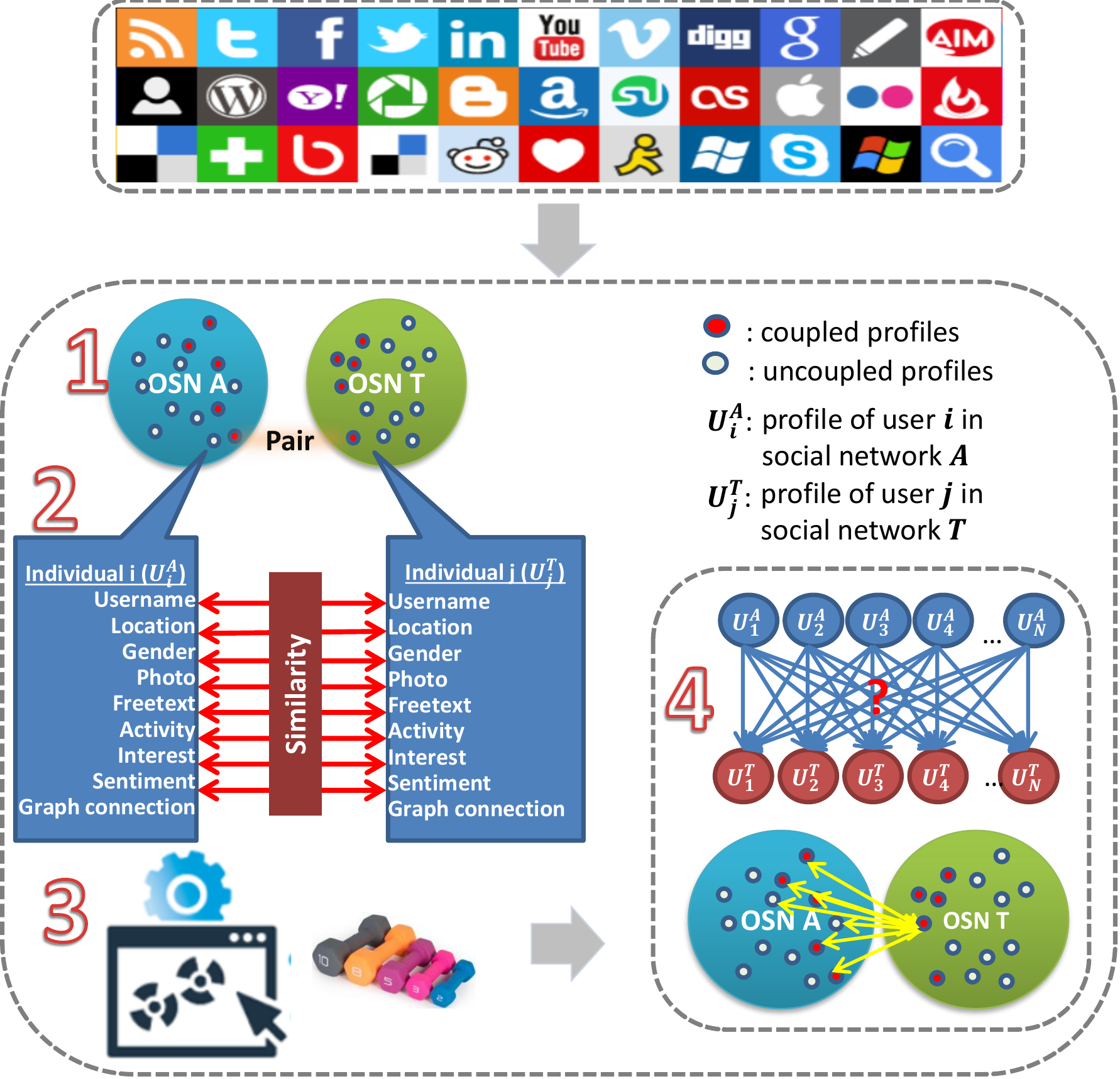}
	\caption{Overview of the proposed profile matching framework in OSNs which consists of 4 main steps: (1) data collection, (2) categorization of attributes and computation of attribute similarities, (3) generating the model, and (4) profile matching.}
	\label{fig:workmodel}
	\vspace{-7pt}
\end{figure}

Let $A$ and $T$ represent the auxiliary and the target OSN, respectively, in which people publicly share attributes such as date of birth, gender, and location. Profile of a user $i$ in either $A$ or $T$ is represented as $U_i^k$, where $k\in\{A,T\}$. In this work, we focus on the most common attributes that are shared in many OSNs. Thus, we consider the profile of a user $i$ as $U_i^k=\{n_i^k,\ell_i^k,g_i^k,p_i^k,f_i^k,a_i^k,t_i^k,s_i^k,r_i^k\}$, where $n$ denotes the user name, $\ell$ denotes the location, $g$ denotes the gender, $p$ denotes the profile photo, $f$ denotes the freetext provided by the user in the profile description, $a$ denotes the activity patterns of the user in a given OSN (i.e., time instances at which she is active), $t$ denotes the interests of the user (on that particular OSN), $s$ denotes the sentiment profile of the user, and $r$ denotes the (graph) connectivity pattern of the user. As discussed, the main goal of the attacker is to link the profiles between two OSNs. The overview of the proposed framework is shown in Figure~\ref{fig:workmodel}.

In general, our proposed framework is composed of two main parts: (i) Steps~1-3 (in Figure~\ref{fig:workmodel}) constitute model generation and they are the offline steps of the algorithm, (ii) Step~4 is the profile matching part. We give a highlevel description of each step in the following. 

In Step~1, profiles and attributes of a set of users are obtained from both OSNs to construct the training dataset. We denote the set of profiles that are extracted for this purpose from OSNs $A$ and $T$ as $\mathrm{A_t}$ and $\mathrm{T_t}$, respectively. We assume that profiles are selected such that some profiles in $\mathrm{A_t}$ and $\mathrm{T_t}$ belong to the same individuals and some do not (more details on collecting such profiles can be found in Section~\ref{sec:datasetcreate}).\footnote{Such profiles are required to construct the ground-truth for training.} We let set $\mathrm{G}$ include pairs of profiles $(U_i^A,U_j^T)$ from $\mathrm{A_t}$ and $\mathrm{T_t}$ that belong to the same individual (i.e., coupled profiles). Similarly, we let set $\mathrm{I}$ include pairs of profiles $(U_i^A,U_j^T)$ from $\mathrm{A_t}$ and $\mathrm{T_t}$ that belong to different individuals (i.e., uncoupled profiles).

In Step~2, for each pair of users in sets $\mathrm{G}$ and $\mathrm{I}$, we compute the attribute similarity by using the metrics that are discussed in Section~\ref{sec:metrics}. In Step~3, we label the pairs in sets $\mathrm{G}$ and $\mathrm{I}$ and add them to the training dataset. If the pair is in set $\mathrm{G}$, we label the pair as ``1'', otherwise we label it as ``0''. We generate our model using different machine learning techniques such as logistic regression and support vector machine to learn the contribution of each attribute to profile matching (details of this step are discussed in Section~\ref{sec:weights}). In Step~4, the attack type is determined and profiles to be matched are selected, and hence sets $\mathrm{A_e}$ and $\mathrm{T_e}$ are constructed. For simplicity, we assume set $\mathrm{A_e}$ includes $N$ users from OSN $A$ and set $\mathrm{T_e}$ includes $N$ users from OSN $T$.\footnote{Sets $\mathrm{A_e}$ and $\mathrm{T_e}$ do not include any users from sets $\mathrm{A_t}$ and $\mathrm{T_t}$.} Every profile in set $\mathrm{A_e}$ is paired with every profile in set $\mathrm{T_e}$ and the similarity between each pair is computed by using the generated model. In the end, profiles in sets $\mathrm{A_e}$ and $\mathrm{T_e}$ are paired by maximizing similarities using an optimization algorithm as discussed in Section~\ref{sec:hung}. 

\subsection{Categorizing Attributes and Defining Similarity Metrics}\label{sec:metrics}

Once the attributes of the users are extracted from their profiles, they should be categorized so that similarity values of attributes between different users can be computed. In the following, we summarize how we categorize the considered attributes and define their corresponding similarity metrics between a user $i$ in OSN $A$ and a user $j$ in OSN $T$. We refer the reader to~\cite{halimi2017profile} for a detailed description of the similarity metrics.
\begin{itemize}
	\item \noindent\textbf{User name similarity - $S(n^A_i,n^T_j)$:} We use Levenshtein distance~\cite{levenshtein} to calculate the user name similarity.
	\item \noindent\textbf{Location similarity - $S(\ell^A_i,\ell^T_j)$:} We convert the textual location information collected from the users' profiles into coordinates via GoogleMaps API~\cite{googlemaps} and calculate geographic distance.
	\item \noindent\textbf{Gender similarity - $S(g^A_i,g^T_j)$:} If an OSN does not provide the gender information publicly (or does not have such information), we probabilistically infer the possible gender information by using the US social security name database\footnote{US social security name database includes year of birth, gender, and the corresponding name for babies born in the United States.} and look for a profile's name (or user name).
	\item \noindent\textbf{Profile photo similarity - $S(p^A_i,p^T_j)$:} We calculate this via a face recognition tool named OpenFace~\cite{amos2016openface}.
	\item \noindent\textbf{Freetext similarity - $S(f^A_i,f^T_j)$:} Freetext data in an OSN profile can be a short biographical text or an ``about me'' page. In this work, we use NER (named-entity recognition)~\cite{ner} to extract features (location, person, organization, money, percent, date, and time) from the freetext information. To calculate the similarity, we use the cosine similarity between the extracted features from each user.
	\item \noindent\textbf{Activity pattern similarity - $S(a^A_i,a^T_j)$:} Activity pattern similarity is defined as the similarity between observed activity patterns of two profiles (e.g., login or post). Let $a^A_i$ represent a vector including the times of last $|a^A_i|$ activities of user $i$ in OSN $A$. Similarly, $a^T_j$ is a vector including the times of last $|a^T_j|$ activities of user $j$ in OSN $T$. First, we compute the time difference between every entry in $a^A_i$ and $a^T_j$, and then we compute the normalized distance of these pairs to compute the activity pattern similarity. 
	\item \noindent\textbf{Interest similarity - $S(t^A_i,t^T_j)$:} First, we create a topic model using the posts of randomly selected users from both the auxiliary and the target OSNs. To create the topic model we use Latent Dirichlet Allocation (LDA)~\cite{Blei:2003:LDA:944919.944937}. Then, by using the created model, we compute the topic distribution of each post generated by the users and compute the interest similarity from the distance of the topic distributions.
	\item \noindent\textbf{Sentiment similarity - $S(s^A_i,s^T_j)$:} To determine whether the shared text expresses positive or negative sentiment we use sentiment analysis tool of Python NLTK (natural language toolkit) text classification~\cite{toolkit}. This tool returns the probability for positive and negative sentiment in the text. Since users' moods are affected from different factors, it is realistic to assume that they may change by time (e.g., daily). Thus, we compute the daily sentiment profile of each user and the similarity between them.
	\item \noindent\textbf{Graph connectivity similarity - $S(r^A_i,r^T_j)$:} To model the graph connectivity pattern of a user, we follow the same strategy as in~\cite{sharad}. For each user $i$, we define a feature vector $F_i=(c_0, c_1, ..., c_{n-1})$ of length $n$ made up of components of size $b$. Each component contains the number of neighbors that have a \textit{degree} in a particular range, e.g., $c_k$ is the count of neighbors with a degree such that $k\cdot b<degree\le (k+1)\cdot b$. We use the feature vector length as $70$ and bin size as $15$ (as in~\cite{sharad}).
\end{itemize}

\subsection{Generating the Model}\label{sec:weights}

As discussed, we first construct sets $\mathrm{A_t}$ and $\mathrm{T_t}$ for training. Also, set $\mathrm{G}$ includes pairs of profiles $(U_i^A,U_j^T)$ that belong to the same individual and set $\mathrm{I}$ includes pairs of profiles $(U_i^A,U_j^T)$ from $\mathrm{A_t}$ and $\mathrm{T_t}$ that belong to different individuals. We refer to the pairs in $\mathrm{G}$ as ``coupled profiles'' and the ones in $\mathrm{I}$ as ``uncoupled profiles''. We first compute the individual attribute similarities between each pair of coupled and uncoupled profiles in $\mathrm{G}$ and $\mathrm{I}$ using the similarity metrics described in Section~\ref{sec:metrics}. Then, to train (and construct) the model and learn the contribution (or weight) of each attribute, we use two different machine learning techniques: (i) logistic regression and (ii) support vector machine (SVM).

\subsection{Matching Profiles}\label{sec:hung}
As discussed, for profile matching, we consider the users in sets $\mathrm{A_e}$ and $\mathrm{T_e}$ from the auxiliary and the target OSNs. For simplicity, we also assume that both sets include $N$ users.\footnote{The case when the sizes of the OSNs are different can be also handled similarly (by padding one OSN with dummy users to equalize the sizes).} Before the actual profile matching, individual attribute similarities between every profile in $\mathrm{A_e}$ and in $\mathrm{T_e}$ are computed using the similarity metrics described in Section~\ref{sec:metrics}. Then, the general similarity $S(U^{A}_i, U^{T}_j)$  is computed between every user in $\mathrm{A_e}$ and $\mathrm{T_e}$ using the weights determined in Section~\ref{sec:weights}. Let $Z$ be a $N \times N$ similarity matrix that is constructed from the pairwise similarities between the users in $\mathrm{A_e}$ and $\mathrm{T_e}$. Our goal is to obtain a one-to-one matching between the users in $\mathrm{A_e}$ and $\mathrm{T_e}$ that would also maximize the total similarity. To achieve this matching, we use the Hungarian algorithm, a combinatorial optimization algorithm that solves the assignment problem in polynomial time~\cite{hungarian}. The objective function of the Hungarian algorithm can be expressed as below.
\begin{equation*}
	min \sum\limits_{i=1}^N \sum\limits_{j=1}^N {{-Z}_{ij}x_{ij}},
\end{equation*}
where, $Z_{ij}$ represents the similarity between $U^A_i$ and $U^T_j$ (i.e., $S(U^A_i,U^T_j)$). Also, $x_{ij}$ is a binary value, that is, $x_{ij}=1$ if profiles $U^A_i$ and $U^T_j$ are matched as a result of the algorithm, and $x_{ij}=0$ otherwise. After performing the Hungarian algorithm to the $Z$ matrix, we obtain a matching between the users in $\mathrm{A_e}$ and $\mathrm{T_e}$ that maximizes the total similarity. Note that we multiply $Z_{ij}$ values with -1, in order to obtain the maximum similarity (profit). We use the one-to-one match obtained from Hungarian algorithm to quantify the privacy risk of OSN users due to profile matching.

\section{Evaluation}\label{sec:evaluation}
In this section, we evaluate the proposed framework by using real data from four OSNs. We also study the impact of various sets of attributes to profile matching.

\subsection{Evaluation Metrics}

To evaluate our model, we consider two types of profile matching attacks: (i) targeted attack, and (ii) global attack. In targeted attack, the goal of the attacker is to match the anonymous profiles of one or more target individuals from $T$ to their corresponding profiles in $A$. In the global attack, the goal of the attacker is to match all profiles in $\mathrm{A_e}$ to all profiles in $\mathrm{T_e}$. In other words, the goal is to deanonymize all anonymous users in the target OSN (who have accounts in the auxiliary OSN).

In both targeted and global attacks, we use Hungarian algorithm for profile matching between the auxiliary and the target OSN (as discussed in Section~\ref{sec:hung}). Hungarian algorithm provides a one-to-one match between all the users in the auxiliary and the target OSN. However, we cannot expect that all anonymous users in the target OSN to have profiles in the auxiliary OSN (we are only interested in the ones that have profiles in both OSNs). Therefore, some matches provided by the Hungarian algorithm are useless for us. Thus, we define a confidence value and we only consider the correct matches above this value to compute the true positives. For this purpose, we set a ``similarity threshold''. For the evaluation metrics, we use precision, recall, and accuracy. We compute accuracy as the fraction of correctly matched coupled pairs to all coupled pairs regardless of the similarity threshold.

\subsection{Data Collection}\label{sec:datasetcreate}

In the literature there are limited datasets that can be used for profile matching between unstructured OSNs. Thus, to evaluate our proposed framework, we collected two datasets that consist of users from three OSNs (Twitter, Foursquare, and Google+) with several attributes. The most challenging part of data collection was to obtain the ``coupled'' profiles between OSNs that belong to same person in real-life. We also used the Flickr social graph~\cite{Zafarani+Liu:2009} to evaluate our proposed framework on structured OSNs. In the following, we discuss our data collection methodology.

\noindent\textbf{Dataset~1 ($\mathrm{D1}$): Twitter - Foursquare}: To collect the coupled profiles, we used Twitter Streaming API~\cite{twitterstreaming}. When an individual generated a check-in in the Swarm app (a companion app to Foursquare)~\cite{swarm} and published it via Twitter, we connected the corresponding Twitter and Foursquare accounts to each other (as coupled profiles). We then removed such simultaneous posts from the dataset. Furthermore, we also randomly paired uncoupled profiles which are used for training and testing the proposed algorithm. We used Foursquare as our auxiliary OSN ($A$) and Twitter as our target OSN ($T$). $\mathrm{D1}$ consists of $4000$ user profiles in each OSN where $2000$ users have profiles in both OSNs.

\noindent\textbf{Dataset~2 ($\mathrm{D2}$): Google+ - Twitter }: To collect the coupled profiles, we exploited the fact that Google+ allows users to explicitly list their profiles in other social networks on their profile pages. We first visited random Google+ profiles and parsed the URLs to Twitter accounts of the users (if it exists). Then, we extracted information from both user profiles. We used Twitter as our auxiliary OSN ($A$) and Google+ as our target OSN ($T$). Note that Google+ has shut down after our data collection. However, results we show using $\mathrm{D2}$ are still good representatives of profile matching risk for OSNs in which users share similar content as Google+ (e.g., Facebook). $\mathrm{D2}$ consists of $8000$ users in each OSN where $4000$ of them are coupled profiles.

\noindent\textbf{Dataset~3 ($\mathrm{D3}$): Flickr social graph~\cite{Zafarani+Liu:2009}:} We generated both target and auxiliary OSN graphs by sampling one whole graph into two pieces as in~\cite{sharad}. To generate the auxiliary and the target OSN graphs, we used a vertex overlap of $1$ and an edge overlap of $0.9$. $\mathrm{D3}$ consists of $50000$ users. 

To create the LDA model, we randomly sampled a total of $15000$ tweets (from Twitter), tips (from Foursquare), and posts (from Google+) and generated the model by using this data. Then, we apply the model to the posts of the users to find the interest similarity as discussed in Section~\ref{sec:metrics}. Note that there may be missing attributes (that are not published by the users) in the dataset. In such cases, based on the distributions of the similarity values of each attribute between the coupled and uncoupled pairs, we assign a value for the similarity that minimizes both the false positive and false negative probabilities.

\subsection{Evaluation Settings}\label{sec:learning}

In the rest of the paper, we will hold the discussion over a target and auxiliary network as the training process is the same for all datasets. As mentioned, in $\mathrm{D1}$, Twitter is the target network and Foursquare is the auxiliary network. In $\mathrm{D2}$, Google+ is the target network and Twitter is the auxiliary network. In $\mathrm{D3}$, both target and auxiliary network is generated from Flickr. From each dataset, we select $3000$ profile pairs for generating the model. These pairs consist of $1500$ coupled and $1500$ uncoupled profile pairs. To generate the model, we use two different machine learning techniques: (i) logistic regression and (ii) support vector machine. Overall, we conduct three experiments by using different sets of attributes. $\mathrm{Experiment~1}$ and $\mathrm{Experiment~2}$ are conducted on $\mathrm{D1}$ and $\mathrm{D2}$ while $\mathrm{Experiment~3}$ is conducted on $\mathrm{D3}$. 

In our first experiment ($\mathrm{Experiment~1}$), we use all the attributes we extracted from both OSNs for the model generation. We observe that location, user name, and profile photo are the most identifying attributes to determine whether two profiles belong to same individual or not. In the second experiment ($\mathrm{Experiment~2}$), we only consider the weak identifiers such as activity patterns, freetext, interests (that is extracted from users' posts), and sentiment. Note that this scenario can be also used to quantify the risk of profile matching between an OSN and a profile in a forum (in which users typically remain anonymous, and activity patterns, freetext, interests, and sentiment are the only attributes that can be learned about the users). In the third experiment ($\mathrm{Experiment~3}$ we use only the graph connectivity attribute to match user profiles. Using the generated model, we compute the general similarity between profiles $U_i^A$ and $U_j^T$ for both machine learning techniques (i.e., logistic regression and SVM).

After generating the model for each experiment, we select $1000$ users from the auxiliary OSN and $1000$ users from the target OSN to construct sets $\mathrm{A_e}$ and $\mathrm{T_e}$, respectively (for each dataset). Note that none of these users are involved in the training set. Among these profiles, we have $500$ coupled pairs and we evaluate the accuracy of our proposed framework based on the matching between these coupled profiles.

Most previous works build different classifiers to determine whether two user profiles are matching or not~\cite{goga2015reliability,malhotra,sharad,vosecky}. We also compare our proposed framework with the existing profile matching schemes that are based on machine learning algorithms. In general, we refer to such schemes as the ``baseline approach''. In the baseline approach, we only use the strong identifiers such as user name, location, gender, profile photo, and the graph connectivity (if it is present). We use our proposed metrics to compute the individual similarities of these attributes. We use K-nearest neighbor (KNN), decision tree, random forest, and SVM techniques to classify the pairs as coupled or uncoupled. In KNN, a pair is assigned to the most common class among its k-nearest neighbors. A decision tree has a tree like structure in which each internal node represents a ``test'' on a feature, each branch represents the result of the test, and each leaf represents a class label. A random forest consists of a multitude of decision trees at training time and for each new example, it outputs the average of the prediction of each tree. In our experiments, random forest consists of $400$ trees. In SVM model, the training data is represented as points in space and the data of different categories are divided by a clear gap. New examples are mapped into the same space and are classified by checking on which side of the gap they fall. To implement this baseline approach, first, we train the classifiers with the training dataset constructed in Section~\ref{sec:weights} (including only user name, location, gender, and profile photo for $\mathrm{D1}$ and $\mathrm{D2}$; and graph connectivity features for $\mathrm{D3}$). Then, based on the trained model, we classify each new pair by using either KNN, decision tree, random forest, or SVM.

\subsection{Results}\label{sec:results}

In real-life, two OSNs do not contain exactly the same set of users. Thus, first, we evaluate the proposed framework by using a dataset that includes both coupled and uncoupled profiles. For the global attack, we try to match all $N=1000$ profiles in $\mathrm{A_e}$ to $N=1000$ profiles in $\mathrm{T_e}$. Among these pairs, $500$ of them are coupled profile pairs and $99500$ are uncoupled profile pairs, and hence the goal is to make sure that these $500$ users are matched with high confidence. In targeted attack, we set the number of target individuals to $100$ from $T$. These $100$ coupled profiles for the targeted attack are randomly picked among $500$ coupled pairs in the test dataset. We run the targeted attack $10$ times and get the average of the results. We run $\mathrm{Experiments~1,~2~and~3}$ (introduced in Section~\ref{sec:learning}) for these settings. For each experiment, we report the precision and recall values for the similarity threshold at which the precision and recall curves (almost) intersect. In Table~\ref{table:results}, we present the results obtained for the logistic regression model for $\mathrm{Experiments~1~and~2}$, and in Table~\ref{table:graph}, we present the results of the logistic regression model for $\mathrm{Experiment~3}$. In general, we observe that the precision, recall, and accuracy of the logistic regression model are higher compared to the SVM model. Due to the space constraints, we do not present the details of the results for the SVM model.

\begin{table}[t]
	\caption{Results of the profile matching scheme (both targeted and global) with both coupled and uncoupled profiles by using logistic regression as the machine learning technique. For $\mathrm{Experiments~1~and~2}$, we report the precision and recall values for the similarity threshold at which the precision and recall curves (almost) intersect.}
	\centering
	\resizebox{\textwidth}{!}{
		\begin{tabular}{l|c|c|c|c|c|c|c|c|c|c|c|c|}
			\cline{2-13}
			& \multicolumn{6}{c|}{$\mathrm{D1}$ (Twitter - Foursquare)} & \multicolumn{6}{c|}{$\mathrm{D2}$ (Google+ - Twitter)}\\
			\cline{2-13}
			& \multicolumn{3}{c|}{Global Attack} & \multicolumn{3}{c|}{Targeted Attack} & \multicolumn{3}{c|}{Global Attack} & \multicolumn{3}{c|}{Targeted Attack}\\
			\cline{2-13}
			& Precision & Recall & Accuracy & Precision & Recall & Accuracy & Precision & Recall & Accuracy & Precision & Recall & Accuracy\\
			\hline
			\multicolumn{1}{|l|}{$\mathrm{Experiment~1}$ (with all attributes)} & $0.79$ & $0.79$ & $58.6\%$ & $0.85$ & $0.85$ & $63\%$ & $0.88$ & $0.89$ & $62\%$ & $0.88$ & $0.89$ & $63\%$\\
			\hline
			\multicolumn{1}{|l|}{$\mathrm{Experiment~2}$ (with the weak identifiers)} & $0.004$ & $0.004$ & $0.4\%$ & $\sim0$ & $\sim0$ & $0\%$ & $0.45$ & $0.46$ &$12\%$ &$0.43$ & $0.43$ & $13\%$\\
			\hline
		\end{tabular}
	}
	\label{table:results}
	\vspace{-5pt}
\end{table}

\begin{table}[t]
	\caption{Results of the profile matching scheme (both targeted and global) for $\mathrm{Experiments~3}$ by using logistic regression as the machine learning technique. Precision and recall values are computed with the similarity threshold at which the precision and recall curves (almost) intersect.}
	\centering
	\resizebox{\textwidth}{!}{
		\begin{tabular}{l|c|c|c|c|c|c|c|c|c|c|c|c|}
			\cline{2-13}
			& \multicolumn{12}{c|}{$\mathrm{D3}$ (Flickr Social Graph)}\\
			\cline{2-13}
			& \multicolumn{6}{c|}{$\mathrm{A_e}=1000$, $\mathrm{T_e}=1000$} 
			& \multicolumn{6}{c|}{$\mathrm{A_e}=500$, $\mathrm{T_e}=500$} \\
			\cline{2-13}
			& \multicolumn{3}{c|}{Global Attack} & \multicolumn{3}{c|}{Targeted Attack} 
			& \multicolumn{3}{c|}{Global Attack} & \multicolumn{3}{c|}{Targeted Attack}\\
			\cline{2-13}
			& Precision & Recall & Accuracy & Precision & Recall & Accuracy
			& Precision & Recall & Accuracy & Precision & Recall & Accuracy\\
			\hline
			\multicolumn{1}{|l|}{$\mathrm{Experiment~3}$ (only graph connectivity)} & $0.72$ & $0.92$ & $83.4\%$ & $0.85$ & $0.81$ & $84\%$
			& $0.93$ & $0.88$ & $92\%$ & $0.91$ & $0.93$ & $90\%$\\
			\hline
		\end{tabular}
	}
	\label{table:graph}
	\vspace{-5pt}
\end{table}
\begin{table}[ht!]
	\caption{Results of the profile matching scheme (both targeted and global) for $\mathrm{Experiments~1~and~2}$ with only coupled profiles by using logistic regression as the machine learning technique. Precision and recall values are computed with the similarity threshold at which the precision and recall curves (almost) intersect.}
	\centering
	\resizebox{\textwidth}{!}{
		\begin{tabular}{l|c|c|c|c|c|c|c|c|c|c|c|c|}
			\cline{2-13}
			& \multicolumn{6}{c|}{$\mathrm{D1}$ (Twitter - Foursquare)} & \multicolumn{6}{c|}{$\mathrm{D2}$ (Google+ - Twitter)}\\
			\cline{2-13}
			& \multicolumn{3}{c|}{Global Attack} & \multicolumn{3}{c|}{Targeted Attack} & \multicolumn{3}{c|}{Global Attack} & \multicolumn{3}{c|}{Targeted Attack}\\
			\cline{2-13}
			& Precision & Recall & Accuracy & Precision & Recall & Accuracy & Precision & Recall & Accuracy & Precision & Recall & Accuracy\\
			\hline
			\multicolumn{1}{|l|}{$\mathrm{Experiment~1}$ (with all attributes)} & $0.82$ & $0.83$ & $65.6\%$ & $0.87$ & $0.87$ & $66\%$ & $0.90$ & $0.90$ & $66.2\%$ & $0.92$ & $0.92$ & $72\%$\\
			\hline
			\multicolumn{1}{|l|}{$\mathrm{Experiment~2}$ (with the weak identifiers)} & $\sim0$ & $\sim0$ & $0.4\%$ & $\sim0$ & $\sim0$ & $1\%$ & $0.71$ & $0.69$ & $12.8\%$ & $0.66$ & $0.66$ & $13\%$\\
			\hline
		\end{tabular}
	}
	\label{table:coupled_attack}
	\vspace{-5pt}
\end{table}

\begin{figure}
	\centering
	\begin{subfigure}[Dataset~1]{\includegraphics[scale=0.2]{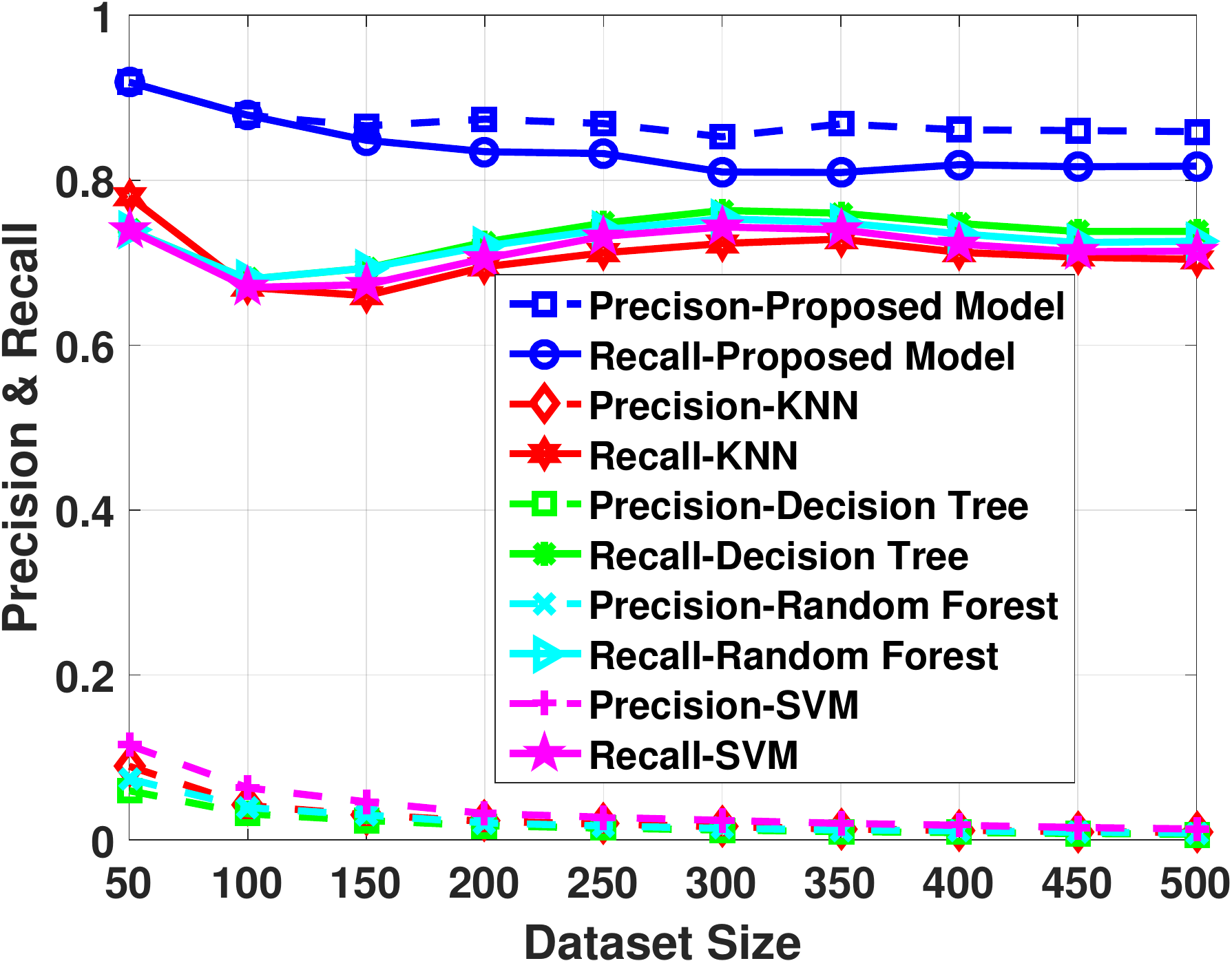}}
	\end{subfigure}\hfill
	\begin{subfigure}[Dataset~2]{\includegraphics[scale=0.2]{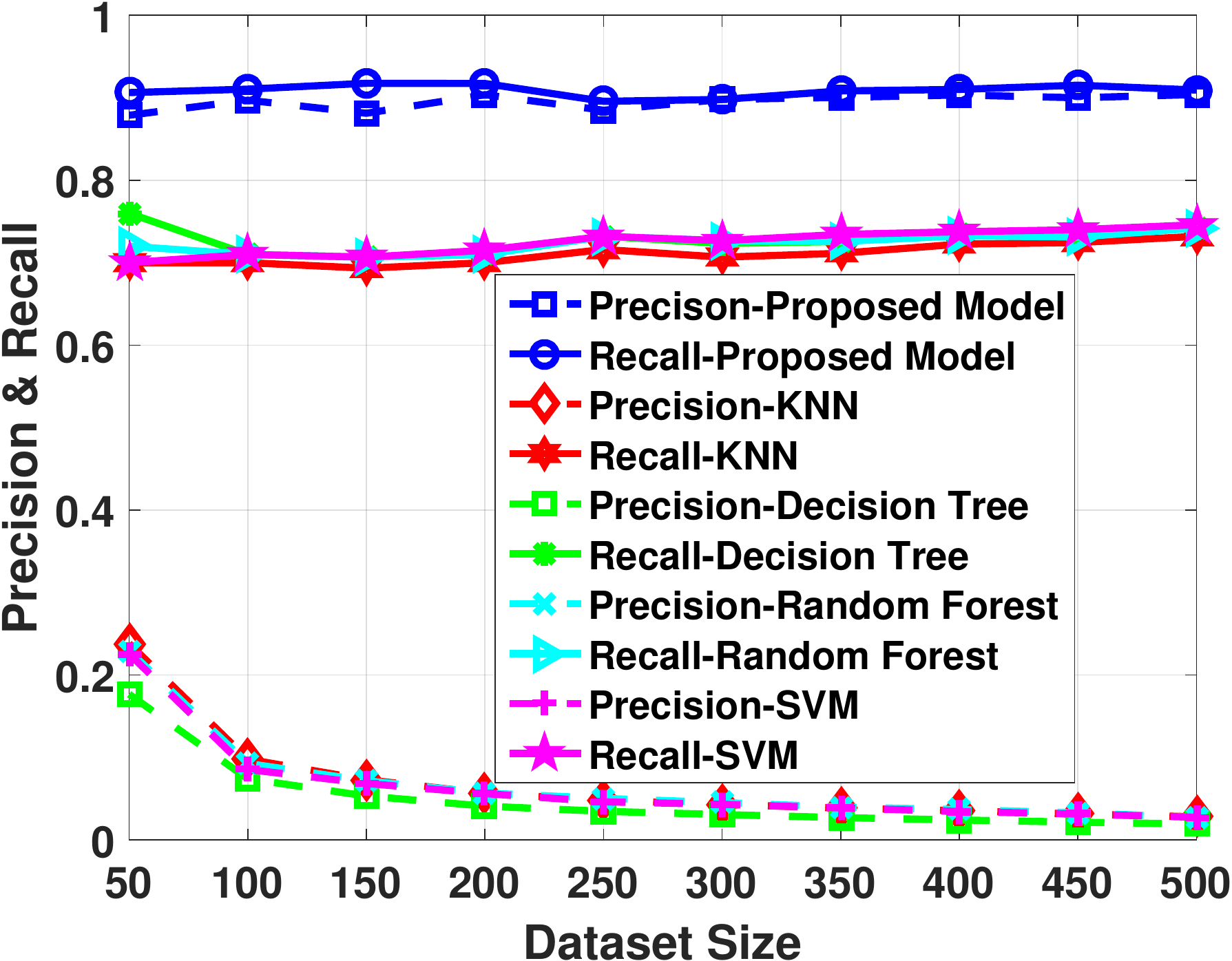}}
	\end{subfigure}\hfill
	\begin{subfigure}[Dataset~3]{\includegraphics[scale=0.2]{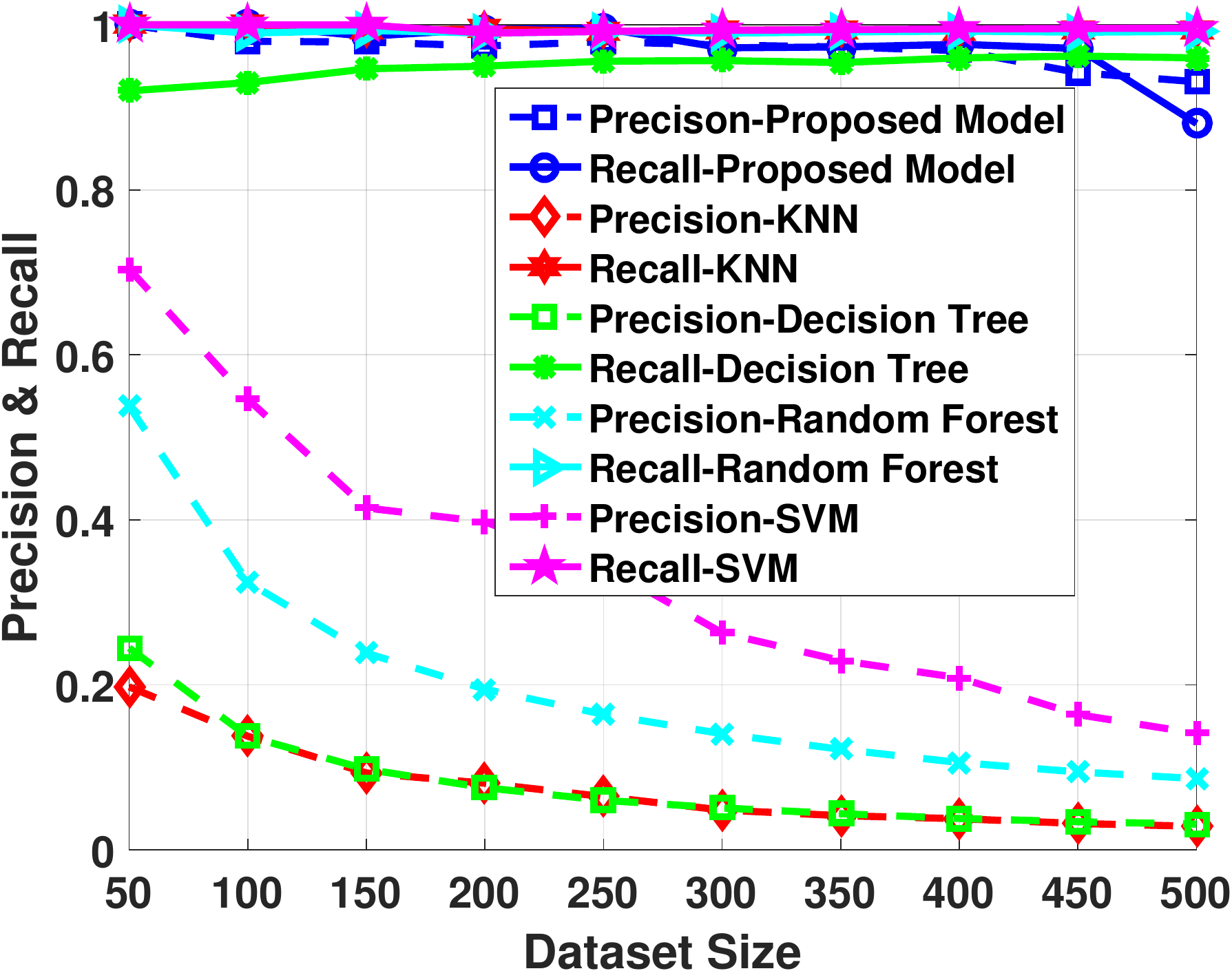}}
	\end{subfigure}\hfill
	\caption{The effect of dataset size to the precision/recall for the global attack in $\mathrm{Experiments~1~and~3}$ with only coupled profiles.}
	\label{fig:coupled_attack}
	\vspace{-10pt}
\end{figure}

In $\mathrm{Experiment~1}$ (in which we use all the attributes), for the global attack, we obtain a precision value of around $0.8$ (for $\mathrm{D1}$) and $0.9$ (for $\mathrm{D2}$) for a similarity threshold of $0.6$. This means that if our proposed framework returns a similarity value that is above $0.6$ for a given profile pair, we can say that the corresponding profiles belong to same individual with a high confidence. Also, overall, we can correctly match $293$ coupled profiles in $\mathrm{D1}$ (with an accuracy of $58.6\%$) and $306$ coupled profiles in $\mathrm{D2}$ (with an accuracy of $62\%$) out of $500$ in global attack. Furthermore, in targeted attack, we obtain a precision value of $0.85$ for $\mathrm{D1}$ and $0.88$ for $\mathrm{D2}$ (for a similarity threshold of 0.6) and overall, we are able to correctly match 63 profiles in both $\mathrm{D1}$ and $\mathrm{D2}$ (out of $100$). Using the same test dataset, we obtain a precision that is close to zero by using the baseline approach (by using KNN, decision tree, random forest and SVM and for both datasets ($\mathrm{D1}$ and $\mathrm{D2}$). This shows that the proposed framework significantly improves the baseline approach while it provides comparable recall value compared to these machine learning techniques (this is further discussed in Figure~\ref{fig:coupled_attack}).

In $\mathrm{Experiment~2}$ (in which we use the weak identifiers), for the global attack, we obtain a precision value of almost $0$ (for $\mathrm{D1}$) and $0.45$ (for $\mathrm{D2}$) and an overall accuracy of $12\%$ for $\mathrm{D2}$. In $\mathrm{Experiment~3}$ (in which we use only the graph connectivity), we obtain a precision value of $0.72$ for $\mathrm{D3}$ in global attack, and we can correctly match $417$ coupled profiles out of $500$ (with an accuracy of $84.7\%$). We further comment on these results in the next section. Overall, the results show that publicly sharing identifying attributes significantly helps profile matching. Furthermore, we show that even the weak identifiers may cause profile matching between the OSN users for some cases.

Next, by only using the $500$ coupled profiles in our test dataset, first we run $\mathrm{Experiments~1,~2~and~3}$ (introduced in Section~\ref{sec:learning}) as before, and then we study the effects of dataset size to profile matching. Thus, for the global attack, we try to match all $N=500$ profiles in $\mathrm{A_e}$ to $N=500$ profiles in $\mathrm{T_e}$ (where there are $500$ coupled and $24500$ uncoupled profile pairs this time) and in targeted attack, we set the number of target individuals to $100$ from $T$ as before. We show the accuracy (i.e., fraction of the correctly matched profiles) and precision/recall values we get from each experiment for the logistic regression model in Tables~\ref{table:graph}~~and~\ref{table:coupled_attack}. As before, in general, we obtain more accurate results for the logistic regression model compared to the SVM model. The precision and recall values reported in the tables are obtained when we set the similarity threshold to the value at which the precision and recall curves (almost) intersect. In practice, the attacker can pick the similarity threshold based on the set of attributes being used for profile matching. In general, we observe that all precision, recall, and accuracy values we obtain for this scenario are higher than the ones reported for the previous scenario (in Table~\ref{table:results}).

Finally, in Figure~\ref{fig:coupled_attack}, we show the precision/recall values of the proposed framework for $\mathrm{Experiments~1~and~3}$ as a function of the dataset size for the global attack and for the logistic regression model. For the proposed framework, we report the precision and recall value for the similarity threshold at which the precision and recall curves almost intersect (as before). In the same figure, we also compare the proposed profile matching scheme with the baseline approach in which we use KNN, decision tree, random forest, and SVM for profile matching as discussed in Section~\ref{sec:learning}. We observe that the precision/recall of the proposed framework does not decrease with increasing dataset size, which shows the scalability of our proposed framework. We also observe that the proposed framework notably provides significantly higher precision values compared to the baseline approach for both $\mathrm{Experiments~1~and~3}$. As shown in Figure~\ref{fig:coupled_attack}, the precision values obtained with the baseline approach are significantly lower than the ones obtained with the proposed framework. This means that the number of false matches (matched profiles that do not belong to the same individual) is high. In order to decrease the number of false matches, one can use a cutoff threshold for the probability returned from the classifier. By doing so, two user profiles are matched only if the probability returned by the classifier is greater than this cutoff threshold. We also compute precision and recall for the baseline approach using different values for such a cutoff threshold and observe that our proposed framework still outperforms the baseline approach. Furthermore, we observe that using such a cutoff threshold causes precision/recall of the baseline approach to decrease with increasing dataset size.
\vspace{-5pt}

\subsection{Discussion}\label{sec:discussion}

In general, for all experiments, we observe that logistic regression provides better results compared to the SVM model. In terms of the variation of the results obtained for different datasets, we observe the followings:
\vspace{-2pt}
\begin{itemize}
	\item Precision, recall, and accuracy obtained from $\mathrm{D2}$ are higher compared to $\mathrm{D1}$. Users share more complete and informative information in Google+ compared to Foursquare. In particular, $\mathrm{Experiment~2}$ shows that Google+ profiles provide more complete information in terms of freetext sharings, activity patterns, and interests of the users. 
	\item $\mathrm{D3}$ (which contains only the network structure of Flickr) achieves a higher accuracy than $\mathrm{D2}$ (and $\mathrm{D1}$) due to the high similarity between the target and the auxiliary OSNs. When the overlap between them is decreased, the accuracy of proposed framework decreases, but still is higher than the one obtained from the baseline approach.
	\item In $\mathrm{D1}$, the weight for the activity pattern is higher than the one for $\mathrm{D2}$ because, some users tend to share about their Foursquare check-ins on their Twitter accounts at close times (there is no such behavior between Google+ and Twitter).
\end{itemize}
\vspace{-2pt}
These observations can also be generalized for other OSNs that share common behavior with the ones that we studied. We also have the following observations in terms of the attributes we used:
\vspace{-2pt}
\begin{itemize}
	\item In $\mathrm{D1}$ and $\mathrm{D2}$, the user name attribute is the most differentiating one compared to others.
	\item Our results show that except user name, other strong identifiers include location, gender, and profile photo. One may claim that users that are matched based on their strong identifiers may not be privacy conscious. That is why in $\mathrm{Experiment~2}$ (in Section~\ref{sec:results}), we remove such strong identifiers and only consider the weak identifiers (activity patterns, freetext, interests, and sentiment) of the users for profile matching. The results show that the contribution of weak identifiers to the profile matching is significantly lower compared to the strong identifiers (as shown in Tables~\ref{table:results}~-~\ref{table:coupled_attack}). However, weak identifiers require more data and analysis. As more posts are collected, we expect that the contribution of the weak identifiers will increase. We will head to this direction in future work. We will also enrich the variety of weak identifiers and collect the graph structure together with the public attributes.
	\item Even though the contribution of the weak identifiers is low, we show that it is still possible to match user profiles by only using them. Note that weak identifiers are hard to be controlled, even for privacy-conscious users. Thus, showing the potential to match user profiles by only using weak identifiers justifies the severity of the matching risk.
\end{itemize}
\vspace{-2pt}

Note that in datasets $\mathrm{D1}$ and $\mathrm{D2}$, users willingly provide links to their social networks, while in $\mathrm{D3}$ auxiliary and anonymized graph are generated from the same graph. We acknowledge that such users might not represent privacy conscious ones. However, it is hard to find groundtruths that represent privacy cautious users. Also, in previous works~\cite{goga2015reliability,zafarani} coupled profiles were obtained in a similar way by using Google+ or about.me, where users provide the links to their social profiles. As future work, we will collect a dataset that contains high number of posts and will focus on profile matching based on weak identifiers.
\vspace{-5pt}

\section{Conclusion and Future Work}\label{sec:conclusion}

In this work, we have proposed a framework for profile matching in online social networks (OSNs) by considering the graphical structure and other attributes of the users. Our results show that by using only public available information, users' profiles in different OSNs can be matched with high precision and accuracy. We have shown how different spectrum of publicly available attributes can be utilized to match user profiles. We have also shown that even a limited number of weak identifiers of the users, such as activity patterns across different OSNs, interest similarities, and freetext similarities may be sufficient for the attacker in some cases. We have shown that the proposed framework significantly improves the baseline approach in terms of precision while providing comparable recall values compared to state of the art machine learning techniques.

As future work, we will work on designing a user interface that informs the users about their privacy risk due to profile matching in real-time (as they share a new content). We will also provide suggestions to the users for alternative sharings (e.g., modify content, share later, or share with more generalized information) in order to reduce the risk. We will work on approximate graph-matching algorithms to improve the efficiency of the proposed framework. We will also extend the work for multiple auxiliary OSNs that may have correlations with each other.

\vspace{5pt}
\noindent\textbf{Acknowledgment.} We thank Volkan K\"u\c{c}\"uk for collecting $\mathrm{D1}$ and $\mathrm{D2}$ and for his help in the initial phases of this work.
\vspace{-7pt}

\bibliographystyle{splncs04}
\bibliography{references}
\end{document}